\documentstyle{article}
\begin{document}
ZU-TH 3/98
\vskip1cm
\hrule
\vskip4mm
\begin{center}
{\bf
accepted for publication to {\it International Journal of Modern Physics D} - 
February 1998
}
\end{center}
\vskip4mm
\hrule
\vskip8mm
\begin{center}
\huge{\bf Primordial magnetic field and spectral distortion of 
cosmic background radiation}
\vskip8mm
\large{\bf Denis Puy$^{1,2}$ and Patrick Peter$^{3}$}
\vskip4mm
$^1$Paul Scherrer Institut\\
Labor fur Astrophysik\\
CH-5232 Villigen PSI (Switzerland)\\
puy@pss027.psi.ch
\vskip2mm
$^2$Institut fur Theoretische Physik\\
Universitat Zurich\\
Winterthurerstrasse, 190\\
CH-8057 Zurich (Switzerland)\\
puy@physik.unizh.ch
\vskip2mm
$^3$Observatoire de Paris-Meudon\\
D\'epartement d'Astrophysique Relativiste et de Cosmologie\\
5, Place Jules Janssen\\
92195 Meudon (France)\\
peter@prunelle.obspm.fr
\end{center}

\section*{Abstract}
The role played by a primordial magnetic field during the
pre-recombination epoch is analysed through the cyclotron
radiation (due to the free electrons) it might produce in
the primordial plasma. We discuss the constraint implied by
the measurement or lack thereof COBE on this primordial field.
\section{Introduction}
The existence of a primordial magnetic field has been the subject of many
discussions. From the theory of turbulence in the primordial plasma developped by
Gamow (1946) and Ozernoy and Chernin (1968), Harrison (1980) proposed a model
for the generation of primordial magnetic field through the cosmic turbulence
theory. Brecher and Blumenthal (1970) have suggested that a possible
alignment of the baryonic and leptonic magnetic moments at early stages of
cosmological evolution could have created a large scale magnetic field.
The parity
non conservation in weak interactions could also provide
a mechanism of magnetic field
generation as proposed by Vilenkin and Leahy (1982). Witten (1985) discussed
a mechanism by which an electromagnetic current could be induced on cosmic
strings, a mechanism that was shown to occur in most theories in which strings
are present (Davis and Peter 1995).
More recently Tajima {\sl et al.} (1992) proposed that
 during the quark-hadron phase transition, density fluctuations
could have generated magnetic fields whose expected order of magnitude
would be much greater than in the case of the cosmic turbulence theory
of Ozernoy.
\\
 It is therefore not completely unrealistic to
consider the existence of a primordial magnetic fields and therefore to 
investigate their cosmological consequences.
\\
The influence of a
primordial magnetic field has been considered in many physical cases
(primordial nucleosynthesis, electromagnetic radiation).
Wassermann (1978), in particular, analyzed the effects of a large-scale
primordial magnetic field on the formation of large scale structures and,
more recently, Coles (1992) pointed out the possible
role of a primordial magnetic
field in the cold dark matter scenario.
\\
A magnetic field can induce a cyclotron radiation due to the acceleration of
the free electrons in the primordial plasma which would yield an energy release.
This radiation can be responsible for the existence of early distortions on
the microwave background spectrum. The aim of this paper is to study the 
influence of this radiation on the distribution of photons. i.e. Kompaneets 
equation (Kompaneets 1957), and the
$\mu$ and $y$ spectral distortions covered by COBE (Mather et al 1993),
 where $\mu$ is the chemical
 potential and $y$ the Compton distortion parameter, and to place a constraint 
on a primordial magnetic field. We introduce
the equations of evolution in Section 2 and, in Section 3, we discuss
the constraint given by COBE on the primordial magnetic field. Then the
limitations of this simple model and its cosmological implications are
discussed as well as the origin of such a primordial magnetic field.
\section{The equations of evolution}
All the equations of evolution are calculated in the case of an Einstein-De
Sitter Universe and we shall assume for numerical calculations a value for the
Hubble constant of $H_o \, = \, 50$ km s$^{-1}$ Mpc$^{-1}$ h.
\subsection{Density}
The density of the matter is known to be proportional to $a^3$ where $a$ is the
factor of scale, so the matter density evolves as
\begin{equation}
n \, = \, n_o (1+z)^3,
\end{equation}
with the present matter density (at $z=0$) being given by
\begin{equation}
n_o \, = \, \frac{3 H_o^2}{8 \pi
G} \, \frac{\Omega_b}{1.4 \, m_H} \, \sim \, 2.04 \, 10^{-7} \, {\rm cm}^{-3}
\end{equation}
where $G$ is the gravitational constant, $\Omega_b =0.1$ the relative
baryonic density with respect to the critical density $\Omega_b= \rho_b /
\rho_c$, with $\rho_c = 3H_o^2/8 \pi G$, and $m_H$ is the hydrogen mass.
\subsection{Temperature}
In the Universe the temperature of the radiation evolves as
\begin{equation}
T_r \, = \, T_o (1+z),
\end{equation}
where $T_o=2.726$ Kelvins is the temperature of the cosmic background radiation (CMBR) at
$z=0$, which is given by the measurement of the COBE FIRAS instrument
(Far InfraRed Absolute Spectrometer, see Mather {\sl et al.} 1993).
\\
Peebles (1968) has shown that the Compton scattering plays a role in the thermal
 balance between matter and radiation. At the beginning of the recombination
period, matter and radiation were very close to thermal equilibrium. 
At the time of recombination, the species $H^+$ and $D^+$ (hydrogen and
deuterium ions) recombine with the free electrons and so considerably
reduce the ionisation fraction. 
Jones and Wyse (1985) have rederived the equations for the ionisation of the
cosmic plasma during and before the recombination period. 
They find that before the recombination period, the medium is practically
fully ionized. For this reason, in what follows, we consider only the 
pre-recombination period, and we take the approximation for the redshift 
$1 + z \sim z$.
\\
Thus matter and radiation are
coupled ($T_r = T_m$), and we have, during the pre-recombination period,
\begin{equation}
T_m \, = \, T_r \, \Rightarrow \, T_m \, = \, T_o \, z.
\end{equation}
\subsection{Magnetic field}
Once the field is imprinted on the charged plasma it will remain there 
because the early Universe is a very efficient conductor. Its conductivity
is inversely proportional to the collision cross section as was shown 
Turner and Widrow (1988). 
\\
A magnetic field frozen in a plasma accelerates all charged particles. In what
follows, we shall only consider the motion of the free electrons, neglecting
all heavy charged particles because of the large mass ratios that enter the
equations. In the dipolar approximation, the Poynting flux gives the
radiative emission $\Psi$ per unit volume (in erg cm$^{-3}$ s$^{-1}$)
\begin{equation}
\Psi \, = \, \frac{2}{3}\, \frac{e^4}{m_e^2 c^3} \,
\frac{\langle v_e^2 \rangle}{c^2} \, B^2 \, n_e
\end{equation}
at the frequency
\begin{equation}
\nu_{cycl} \, = \, \frac{e B}{2 \pi m_e c},
\end{equation}
where $e$ is the electric charge of an electron, $m_e$ his mass, $B$ the 
magnetic field, $n_e$ the density of electrons close to the matter density. 
The electron velocity $v_e$ is supposed to be
of purely thermal origin, i.e.,
\begin{equation}
 \langle v_e^2\rangle  = \frac{3 k T_m}{m_e},
\end{equation}
where $k$ is the Bolzmann constant. Moreover the medium is practically fully 
ionized, so we will assume $n_e \sim n_{protons} \sim n$. Therefore we can write
\begin{equation}
\Psi \, = \, \frac{2 e^4 \,k T_o n_o  B^2}{m_e^3 c^5} (1+z)^4
\end{equation}
for emissions occuring at redshift $z$.
\\
The evolution of the magnetic field is due to the Cyclotron effect and the 
expansion. We now explore both effects to show that the most important reason 
why a primordial magnetic field should decrease with time even though the 
Universe is roughly a perfect conductor in the expansion.
\vskip3mm
\subsection{Cyclotron effect}
Let us introduce the magnetical energy density 
\begin{equation}
\epsilon_{mag} \, = \, 
\frac{B^2}{8 \pi \mu _o}
\end{equation}
and the Cyclotron energy density $\epsilon_{cycl}$ which the 
magnetic field is responsible for. Energy conservation requires  
\begin{equation}
\frac{d\epsilon_{mag}}{dt} \, = \, \frac{d \epsilon_{cycl}}{dt},
\end{equation}
which leads to 
\begin{equation}
\frac{d}{dt} \, \Bigl( \frac{B^2}{8 \pi \mu_o} \Bigr) \, = \, - \Psi
\end{equation}
where $\Psi$ is the radiative emission calculated above. We therefore obtain 
\begin{equation}
\frac{dB}{dt} \, = \,- \frac{8 \pi e^4 \mu_o}{m_e^3 c^5} k T_{o} n_o B (1+z)^4
\end{equation}
In the case of the radiation dominated Universe, we have 
\begin{equation}
\frac{dz}{dt} \, = \, 
-(1+z)^3 H_o^r
\end{equation}
where $H_o^r$ is a constant (in the radiation dominated Universe) given 
by $H_o^r = 8 \pi G \rho_{o,r}/3$ with the density of radiation 
$\rho_{o,r}$ at the redshift $z=0$; thus
\begin{equation}
\frac{dB}{dz} \, = \, 2 \omega (1+z) B
\end{equation}
where
\begin{equation}
\omega \, = \, \frac{4\pi e^4 k T_o n_o \mu _o}{m_e^3 c^5 H_o^r} \, \sim \, 
2.1 \, 10^{-9}
\end{equation}
\subsection{Expansion effect}
The expansion contribution is calculated within the {\it flux conservation} 
framework (Turner and Widrow 1988) which gives here

\begin{equation}
B.a^2 \, = \, constant
\end{equation}
where $a$ is the scale parameter, this leads to
\begin{equation}
\frac{dB}{dz} \, = \, \frac{2B}{1+z}.
\end{equation}
\vskip3mm
Eqs. (14) and (17) combined yield the actual evolution of the magnetic 
field, taking into account both effects, namely
\begin{equation}
\frac{dB}{dz} \, = \, 2 \omega (1+z) B +\frac{2B}{1+z}
\end{equation}
whose solution is easily found as 
\begin{equation}
B\, = \, B_o (1+z)^2 e^{2\omega z + \omega z^2}
\end{equation}
where $B_o$ is the magnetic field at $z=0$. In our case where the distribution 
of photons is closed to a Bose-Einstein distribution we have $z<10^{6.4}$ (see 
Danese \& de Zotti 1977) thus exp$(2 \omega z + \omega z^2) \sim 1$ which lead to neglect 
the relaxation process and the evolution $B \sim B_o \, z^2$. 
Finally for the magnetic flux, we 
have 
\begin{equation}
\Psi \, = \, 
{\cal A} \, B_o ^2 z^8
\end{equation}
with ${\cal A}= \frac{\omega}{2 \pi} \frac{H_o^r}{\mu _o} 
\, \sim \, 4.5 \, 10^{-30}$ c.g.s.

\section{Evolution of the photons distribution}
The Kompaneets equation caracterizes the evolution of the photon distribution 
$\eta$, or the dimensionless {\it occupation number} for the photon gas in 
equilibrium. This equation is, in fact, a simplified Boltzmann equation 
(Gould 1972), and assumes a uniform and isotropic photon gas, which can be 
written as
\begin{equation}
\frac{\partial \eta}{\partial t} \, = \, \Sigma \Lambda_i \, + \, 
\Sigma \Gamma_i
\end{equation}
where the terms $\Lambda_i$ characterize the interactions (collisions) 
involving photons, and $\Gamma_i$ represent sources and sinks of photons. These terms 
depend, in general, of the background.
\\
In our case, namely the pre-recombination period, the dominant 
terms are expected to be Compton scattering $\Lambda_{compt}$ and Bremsstrahlung 
$\Gamma_{brem}$ (Danese and De Zotti 1977). 
However, with a magnetic field present, 
Cyclotron processes can participate, so in the evolution 
equation (21), we must add a cyclotron production term $\Xi_{cycl}$
\begin{equation}
\frac{\partial \eta}{\partial t} \, = \, \Lambda_{compt} \, + \, 
\Gamma_{brem} \, + \, \Xi_{cycl}
\end{equation}
We shall now recall and evaluate the three terms in turn
\subsection{Compton scattering $\Lambda_{compt}$}
The details of the calculation are given in Danese and De Zotti (1977), 
and it is found that 
\begin{equation}
\Lambda_{compt} \, = \, a_c . \frac{1}{x_e}.\frac{\partial}{\partial x_e}
\Bigl[x_e^4 \Bigl(\frac{\partial \eta}{\partial x_e} + \eta (1 + \eta) 
\Bigr) \Bigr]
\end{equation}
where 
\begin{equation}
x_e \, = \, \frac{h \nu}{k T_r}
\end{equation}
and
\begin{equation}
a_c^{-1} \, = \, \Bigl( n \xi_e \sigma_T c \frac{k T_m}{m_e c^2} \Bigr)^{-1}
\, \sim \, 3.7 \, 10^{28} (1+z)^{-4}
\end{equation}
where $\nu$ the frequency of photon, and $\sigma_T$ is the Thomson cross 
section.
\subsection{Bremsstrahlung processes $\Gamma_{brem}$}
The expression of the bremsstrahlung effect is given by (Danese and De Zotti 
1977)
\begin{equation}
\Gamma_{brem} \, = \, K_o g(x_e) \frac{e^{-x_e}}{x_e^3} 
\Bigl[ 1 + \eta (1 -e^{x_e})\Bigr]
\end{equation}
where
\begin{equation}
K_o \, = \, 2.46 \, 10^{-25} \, (1+z)^{5/2}
\end{equation}
and $g(x_e)$ is the Gaunt factor. Let us note that in our case where 
$x_e < 1$, the Gaunt factor may be estimated through the Born approximation 
\begin{equation}
g(x_e) \, = \, \frac{\sqrt{3}}{\pi} \, ln(2.25/x_e)
\end{equation}
\subsection{Cyclotron production $\Xi_{cycl}$}
The emitted power due to cyclotron effect is given by (in erg s$^{-1}$)
\begin{equation}
{\cal P}_{cycl} \, = \, 
\frac{\Psi}{n} \, = \, X B_o ^2 (1+z)^5 
\end{equation}
with 
$$
X \, = \, \frac{\omega H_o^r}{2 \pi \mu _o n_o}\, \sim \, 2.2 \, 10^{-24} \, 
{\rm erg s}^{-1} \, {\rm Gauss}^{-2}
$$
 Finally we deduce the cyclotron contribution to the photons distribution $\eta$
\begin{equation}
\Xi _{cycl} \, = \, {\cal P}_{cycl} \, \delta (\epsilon - \epsilon _{cycl} )
\, = \, X z^5 B_o ^2 \delta ( \epsilon - \epsilon _{cycl} ) 
\end{equation}
where the energy $\epsilon = h \nu$ and the cyclotron energy 
\begin{equation}
\epsilon _{cycl} \, = \, h \nu_{cycl} \, = \, \frac{h e}{2 \pi m_e c} B_o z^2
\end{equation}
\subsection{Evolution of the photons distribution}
Danese and De 
Zotti (1977) and Salati (1992) have discussed the influence of the Compton 
scattering and Bremsstrahlung on the distribution of photons. 
Thomson collisions (Compton scattering) 
redistribute the additional energy, and the spectrum approach a Bose-Einstein 
distribution with chemical potential $\mu$. 
\begin{equation}
\eta(x_e) \, = \, \frac{1}{e^{x_e + \mu} -1}
\end{equation}
The radiation 
spectrum is of Bose-Einstein type with a chemical potential $\mu$, the 
simultaneous 
action of the Bremsstrahlung leads to a Planck distribution (i.e. $\mu 
\rightarrow 0$). Thus a production process of photons work to diminish the potential 
$\mu$, which is characteristic to the energy released, Danese and de Zotti 
give 
\[ 
 \mu \, \sim \, 
\left\{ \begin{array}{ll}
3 \ln [ 0.85 (1 +\frac{\Delta \epsilon}{\epsilon_r}) ] & 
\mbox{if $\mu \gg 1$} \\
1.4 \, \frac{\Delta \epsilon}{\epsilon_r} & \mbox{if $\mu \ll 1$}
\end{array}
\right. 
\]
Nevertheless the cyclotron emission is produced 
only at the cyclotron frequency. 
The Bose-Einstein distribution are not altered by the Cyclotron process, 
we have no effects on the Kompaneets equation. The Cyclotron process will 
induce only spectral distortions due to electromagnetic energy emission in the 
CMBR. 
\section{Spectral distortions due to cyclotron process}
The distortions depend on the time at which the energy is released into 
the primordial plasma. Sunyaev and Zeldovich (1970) discussed the problem of the
interaction of matter and radiation and heating of the primeval plasma. More
recently Salati (1992) has presented the possibilities of scenario of the
spectral distortions of the microwave background radiation, which depend of 
the redshift.
\\
\\
$z > z_P$, redshift at which the double Compton reaction on the electron $e$:
$$
e + h\nu \, \longmapsto \, e + h\nu _1 + h \nu _2
$$
which produces a double emission of photons $h \nu _1$ and $h \nu _2$, becomes
inefficient. Because of this mechanism, any spectral distortion is
smoothed and CMBR is not affected,
The photon distribution is always a planckian distribution.
The value of $z_P=10^{6.4}$ was calculated by Danese and De Zotti (1977).
\\
\\
$z_{BE}< z < z_{P}$,
\\
In this range, the radiation spectrum relaxes towards a
Bose-Einstein (BE) distribution faster than the expansion time whereas the
double Compton emission is slower (see for instance Danese and De Zotti 1977,
Bond 1988, Salati 1992). The average photon energy becomes larger than for a
Planckian spectrum but in thermal equilibrium, which means a nonzero chemical
potentiel $\mu$. Danese and De Zotti (1977) have given the
lower limit $z_{BE} = 10^{4.7}$ and shown that the chemical potential can be
related with the energy release. 
\\
In our context, the energy release is a
magnetic contribution due to the cyclotron radiation, so we have
$$
\mu \, \sim \, 1.4\, \int_{t_{P}}^{t_{BE}} \, \frac{\Psi dt}{a_{bb} T_r^4}
\, = \, 
1.4 \, \int_{z_{BE}}^{z_{P}} \, \frac{\Psi}{a_{bb} T_r^4}.\frac{dz}{H_o^r 
(1+z)^3}
$$
where $t_{BE}$ and $t_P$ represent the age of the Universe respectively at
$z_{BE}$ and $z_P$, so
\begin{equation}
\mu \, \sim \,
\frac{1.4 \omega B_o^2}{2 \pi a_{bb} T_o^4 \mu_o}\, \int_{z_{BE}}^{z_{P}} \, 
(1+z) \,dz
\end{equation}
$z_{rec} < z < z_{BE}$, where $z_{rec}$ corresponds to the redshift of the
recombination of hydrogen.
\\
The last measurement of the cosmic microwave background spectrum derived from
the FIRAS instrument on the COBE satellite give stringent limits on energy
release, the dimensionless cosmological distortion parameter being limited to
$\mu < \, \mu_{FIRAS}= 3.3 \, 10^{-4}$. 
\\
Thus we obtain a constraint on the primordial magnetic field
\begin{equation}
B_o \, < \, 
\sqrt{\frac{4 \pi a_{bb} T_o^4 \mu_{FIRAS} \mu_o}{1.4 \omega z_P^2}}
\end{equation}
i.e. numerically $B_o \, < \, 3.44 \, 10^{-10}$ Gauss
\\
In this range the chemical potential relaxes towards zero ($\mu \rightarrow 0$)
, while the photons and electrons still exchange energy through the Compton
diffusion. The energy release yields a distortion on the Compton distortion
parameter $y$. Danese and De Zotti (1977) have shown that the energy release
is related to $y$, so we can write
\begin{equation}
y \, = \, \frac{1}{4} \, \int_{t_{BE}}^{t_{rec}}  \frac{\Psi dt}{a_{bb} T_r^4}
\, = \, \frac{\omega B_o^2}{8 \pi a_{bb} T_o^4 \mu_o} \, 
\int_{z_{rec}}^{z_{BE}} \, (1+z) \, dz.
\end{equation}
The measure done by FIRAS gives $y < y_{FIRAS}=2.5 \, 10^{-5}$, which again can be
expressed as an upper limit for the magnetic field at $z=0$, namely:
\begin{equation}
B_o \, < \, \sqrt{\frac{16 \pi a_{bb} T_o^4 y_{FIRAS}\mu_o}{\omega z_{BE}^2}}
\end{equation}
which gives $B_o \, < \, 1.12 \, 10^{-8}$ Gauss. 
So the best upper limit is given by the Bose Einstein distortion giving 
$B_o \, < \, 3.44 \, 10^{-10}$ Gauss
\\
This value is more restrictive 
than the constraint on a relic magnetic field
(cluster-sized and unidirectional) such as was recently measured in the Coma
cluster halo by Kim {\sl et al.} (1990) who obtained an upper limit of
$B < 2\times 10^{-8} \, {\rm Gauss}$.
\\
The origin of a magnetic field having these orders of magnitude
at $z=0$ is far from obvious. One interesting possibility,
as pointed out by Thompson (1990), relies on the existence of
superconducting currents trapped in cosmic strings. Such currents,
spacelike or timelike, would be carried by the strings at velocities
approaching that of light at the time they are formed\footnote{in most
models at the Grand Unified scale, but also, as proposed by Peter
(1992-a), following a suggestion by Carter (1990), possibly at much
lower energy scale, in practice close to that of the electroweak
symmetry breaking.}. These strings, whose precise motion in the
primordial cosmological plasma has not yet been investigated
in detail, would carry at least (in the low energy case), roughly
$10^6$~A and 1~C~.~m$^{-1}$ (Peter, 1992). Signore and S\`anchez (1991)
discussed the millimiter and radioastronomical constraints on the cosmological
evolution of superconducting strings, but they did not consider the possible
magnetic field produced. It is clear that the
resulting electromagnetic effects deserve further investigation, including
in particular the precise determination of the remnant magnetic field
as well as its coherence length.
\section*{Acknowledgements.} The authors gratefully acknowledge interesting
suggestions to Monique Signore, Leonid
Ozernoy, Luigi Danese and Gianfranco de Zotti for stimulating criticisms and 
fruitful comments. Part of the work of D. Puy has been conducted under the 
auspices of the {\it D$^r$ Tomalla Foundation} and the Swiss National
Science Foundation.

\end{document}